\documentclass{article}
\usepackage{amssymb}

\begin{document}

\title{Fractional networks, the new structure}
\author{\textit{R. Vilela Mendes\thanks{%
rvmendes@fc.ul.pt, rvilela.mendes@gmail.com, http://label2.ist.utl.pt/vilela/%
}} \\
CMAFCIO, Universidade de Lisboa,\\
Faculdade de Ci\^{e}ncias C6, 1749-016 Lisboa, Portugal}
\date{ }
\maketitle

\begin{abstract}
Real world networks have, for a long time, been modelled by scale-free
networks, which have many sparsely connected nodes and a few highly
connected ones (the hubs). However, both in society and in biology, a new
structure must be acknowledged, the fractional networks. These networks are
characterized by the existence of very many long-range connections, display
superdiffusion, L\'{e}vy flights and robustness properties different from
the scale-free networks.
\end{abstract}

\section{Introduction}

The scale-free network \cite{Barabasi} has been for some time the preferred
paradigm for modeling real world networks in society, biology, etc.
Characterized by asymptotic power-law degree distribution, these networks
have many sparsely connected nodes and a few highly connected ones (the 
\textbf{hubs}). The hubs are the critical nodes to address (or protect) in a
network, because they control the robustness of the network and the
diffusion of information. This is\ a fact well known by politicians,
advertisement agencies and hackers. The importance of hubs has been known
for a long time, even at the time of the Inquisition \cite{Ormerod}. Several
mechanisms, preferential attachment or fitness for example, have been
proposed to explain the formation of this network structure. Many networks
have been reported to be scale-free although careful statistical analysis
has questioned others \cite{Shalizi}.

A feature that has recently emerged in some social networks (see for example 
\cite{Hogan} \cite{Romantic} \cite{Carvalho}) is the existence of very many
long-range connections, rather than hubs. In a sense society imitates Nature
because also brain network phenomena, for example, have been shown to be
dependent on many long-range connections \cite{Markov} \cite{Modha} \cite%
{Knosche} \cite{Barttfeld}. Also the human mobility network has long range
connections of great relevance in epidemiology \cite{Gustafson}. It is to be
expected that the existence of a sufficient number of long range connections
in a network would lead to new phenomena and have a strong effect on the
propagation of information.

Some authors have already studied dynamics on networks involving jumps over
many links leading to fractional diffusion \cite{Riascos}. What one wants to
emphasize here is that fractional diffusion and other phenomena emerge
naturally as a structural property in networks with long range connections.
Hence these networks should be classified as a new structure.

\section{Long-range connections: L\'{e}vy flights and superdiffusion}

\subsection{The Laplacian and Random Walk matrices}

The Laplacian and the Random Walk matrices are the main
tools in the study of dynamical properties in the network. The Laplacian
matrix is%
\begin{equation}
L=G-A  \label{2.1}
\end{equation}%
$G$ being the degree matrix ($G_{ij}=\delta _{ij}\times $ number of
connections of node $i$) and $A$ the adjacency matrix ($A_{ij}=1$ if $i$ and 
$j$ are connected, $A_{ij}=0$ otherwise).

The Random-Walk matrix is%
\begin{equation}
R=G^{-1}A  \label{2.2}
\end{equation}%
then, $R_{ii}=0$, $R_{ij}=\frac{A_{ij}}{\textnormal{degree}\left( i\right) }$ $%
\left( i\neq j\right) $\ if $i$ and $j$ are connected and $R_{ij}=0$
otherwise.

For a node $i$ connected to two other nodes $i+1$ and $i-1$ the action of
the Laplacian matrix on a vector $\left( 
\begin{array}{c}
\vdots \\ 
\psi \left( i-1\right) \\ 
\psi \left( i\right) \\ 
\psi \left( i+1\right) \\ 
\vdots%
\end{array}%
\right) $ leads to $-\psi \left( i-1\right) +2\psi \left( i\right) -\psi
\left( i-1\right) $, which is a discrete version of $-d^{2}$ (minus the
second derivative). Let now $\psi \left( i\right) $ for each node $i$ be the
intensity of some function $\psi $ across the network. It is reasonable to
think that $\psi $ diffuses from $i$ to $j$ proportional to $\psi \left(
i\right) -\psi \left( j\right) $ if $i$ and $j$ are connected. Then,%
\begin{equation}
\frac{d\psi \left( i\right) }{dt}=-k\sum_{j}A_{ij}\left( \psi \left(
i\right) -\psi \left( j\right) \right) =-k\left( \psi \left( i\right)
\sum_{j}A_{ij}-\sum_{j}A_{ij}\psi \left( j\right) \right)  \label{2.3}
\end{equation}%
which in matrix form is%
\begin{equation}
\frac{d\psi }{dt}+kL\psi =0  \label{2.4}
\end{equation}%
a heat-like equation. Therefore the Laplacian matrix controls the diffusion
of quantities in the network.

On the other hand, the $R$ matrix controls the random motion of a walker on
the network. The probability for a random walker to be at the node $i$ at
time $t$ given that at time $t-1$ was at the node $j$ is%
\begin{equation}
p_{i}\left( t\right) =\sum_{j}\frac{A_{ij}}{\textnormal{degree}\left( j\right) }%
p_{j}\left( t-1\right)  \label{2.5}
\end{equation}%
or, in matrix form%
\begin{equation}
p\left( t\right) =G^{-1}Ap\left( t-1\right)  \label{2.6}
\end{equation}

\subsection{A network with power-law connection probability}

Let the network $N$ be embedded into an Euclidean network where distances
may be defined. In the actual network the distances might mean geographical
distances, separation of communities, functional separation as in a brain
network, etc.

In the network, with $A_{ij}=0$ or $1$, let the probability of establishment
of a link at distance $d$ be proportional to a power of the distance%
\begin{equation}
P_{ij}=cd_{ij}^{-\gamma }\textnormal{\hspace{2cm}with }\gamma \leq 3  \label{3.1}
\end{equation}%
To find the nature of the diffusion in such a network, consider a block
renormalized network $N^{\ast }$ where each set of $q$ nearby nodes of $N$
are mapped to a node of the $N^{\ast }$ network. Therefore in the $N^{\ast }$
network the connections are%
\begin{equation}
A_{ij}^{\ast }\simeq cqd_{ij}^{-\gamma }  \label{3.2}
\end{equation}%
Then denoting by $L^{\ast }$ and $G^{\ast }$ the Laplacian and degree
matrices of the $N^{\ast }$ network 
\begin{equation}
L^{\ast }\psi \left( i\right) =G_{ii}^{\ast }\psi \left( i\right)
-cq\sum_{j\neq i}d_{ij}^{-\gamma }\psi \left( j\right)  \label{3.3}
\end{equation}%
What kind of diffusion does the Laplacian matrix $L^{\ast }=G^{\ast
}-A^{\ast }$ imply for the network $N^{\ast }$?. Consider a fractional
diffusion equation%
\begin{equation}
\frac{d\psi }{dt}=-kD^{\beta }\psi  \label{3.4}
\end{equation}%
Using a symmetrized Gr\"{u}nwald-Letnikov representation of the fractional
derivative $\left( a<x<b\right) $ (see for example \cite{Mainardi})%
\begin{eqnarray}
D^{\beta }\psi \left( x\right) &=&\frac{1}{2}\lim_{h\rightarrow 0}\frac{1}{h}%
\left\{ \sum_{n=0}^{\left[ \frac{x-a}{h}\right] }\left( -1\right) ^{n}\left( 
\begin{array}{c}
\beta \\ 
n%
\end{array}%
\right) \psi \left( x-nh\right) \right.  \nonumber \\
&&\left. +\sum_{n=0}^{\left[ \frac{b-x}{h}\right] }\left( -1\right)
^{n}\left( 
\begin{array}{c}
\beta \\ 
n%
\end{array}%
\right) \psi \left( x+nh\right) \right\}  \label{3.5}
\end{eqnarray}%
with coefficients%
\begin{equation}
\left\vert \left( 
\begin{array}{c}
\beta \\ 
n%
\end{array}%
\right) \right\vert =\frac{\Gamma \left( \beta +1\right) \left\vert \sin
\left( \pi \beta \right) \right\vert }{\pi }\frac{\Gamma \left( n-\beta
\right) }{\Gamma \left( n+1\right) }\backsim _{n>>}\frac{\Gamma \left( \beta
+1\right) \left\vert \sin \left( \pi \beta \right) \right\vert }{\pi }%
n^{-\left( \beta +1\right) }  \label{3.6}
\end{equation}%
and $sign\left( 
\begin{array}{c}
\beta \\ 
n%
\end{array}%
\right) =\left( -1\right) ^{n+1}$.

Comparing Eq.(\ref{3.5}) with the expression (\ref{3.3}) for $L^{\ast }\psi
\left( i\right) $, the conclusion is that diffusion in the $N^{\ast }$
network is fractional diffusion of exponent $\beta =\gamma -1$. $\beta =2$
would be normal diffusion, all $\beta <2$ correspond to superdiffusions.
That is, the spreading in time $\left\langle x^{2}\left( t\right)
\right\rangle $ of a distribution localized at $x=0$ at $t=0$ is 
\[
\left\langle x^{2}\left( t\right) \right\rangle \backsim t^{\frac{2}{\beta }}
\]

On the other hand, analyzing the structure of the random walks controlled by 
$G^{-1}A$ (Eq.\ref{2.6}) the conclusion is that whereas for normal diffusion
the jumps are of one step, for $\gamma <3$ arbitrarily large large jumps
occur with a power law (L\'{e}vy flights).

\section{Conclusions}

1. The first general conclusion is that in these networks both \ mobility
and diffusion of information occur at a very fast rate. Therefore it may
considered as a new structure distinct from SF networks, a Fractional
Network (FR). The new structure has wide implications for the control of the
networks.

2. In a SF network, the hubs are both the strength and the weakness of the
network. They insure global connectivity even if a large number of links are
destroyed. But when directly targeted the network is deeply affected
(targeted structural weakness). In a SF network propagation of ideas,
opinions, fads (memes) are most effective if introduced to the hubs. However
fast global establishment of a trend requires its introduction at many hubs.

3. A FR network is structurally very stable and resilient to attack. It is
pointless or too expensive to disrupt the network. The hubs are no longer
the controllers. The network itself is the HUB. Superdiffusion is both the
strength and the weakness of the network. Well crafted memes propagate very
fast. But also do counter-memes. In SF networks the memes are most
efficiently introduced at the hubs. Here they might be introduced anywhere.

\end{document}